\documentclass{article}
\usepackage{enumitem}
\usepackage[top=1cm, bottom=2cm, left=2cm, right=2cm]{geometry}
\usepackage{natbib}
\usepackage{graphicx}
\usepackage{xr}
\usepackage[table,xcdraw]{xcolor}
\usepackage{hyperref}

\usepackage{amsmath}
\usepackage{authblk}
\usepackage{booktabs, makecell, array, tabularx}
\usepackage[T1]{fontenc}
\usepackage[utf8]{inputenc}
\usepackage{fancyhdr}
\usepackage{tcolorbox, tikz}

\title{Automated Extraction of Multicomponent Alloy Data Using Large Language Models for Sustainable Design}

\author[1]{Aravindan Kamatchi Sundaram}
\author[1]{Mohit Chakraborty}
\author[1]{Sai Mani Kumar Devathi}
\author[1]{B. Pabitramohan Prusty}
\author[1,2,*]{Rohit Batra}

\affil[1]{Department of Metallurgical and Materials Engineering, Indian Institute of Technology Madras, Chennai 600036, India}
\affil[2]{Center for Atomistic Modelling and Materials Design, IIT Madras, Chennai 600036, India}
\affil[*]{Author to whom correspondence should be addressed.}

\begin{document}
\newcolumntype{C}{>{\centering\arraybackslash}X}
\maketitle

\begin{abstract}

The design of sustainable materials requires access to materials performance and sustainability data from literature corpus in an organized, structured and automated manner. 
Natural language processing approaches, particularly large language models (LLMs), have been explored for materials data extraction from the literature, yet often suffer from limited accuracy or narrow scope. In this work, an LLM-based pipeline is developed to accurately extract alloy-related information from both textual descriptions and tabular data across the literature on high-entropy (or multicomponent) alloys (HEA). Specifically two databases with 37,711 and 148,069 entries respectively are retrieved; one from the literature text, consisting of alloy composition, processing conditions, characterization methods, and reported properties, and other from the literature tables, consisting of property names, values, and units. The pipeline enhances materials-domain sensitivity through prompt engineering and retrieval-augmented generation and achieves F1-scores of $\sim$0.83 for textual extraction and $\sim$0.88 for tabular extraction, surpassing or matching existing approaches. Application of the pipeline to over 10,000 articles yields the largest publicly available multicomponent alloy database and reveals compositional and processing-property trends. The database is further employed for sustainability-aware materials selection in three application domains, i.e., lightweighting, soft magnetic, and corrosion-resistant, identifying multicomponent alloy candidates with more sustainable production while maintaining or exceeding benchmark performance. The pipeline developed can be easily generalized to other class of materials, and assist in development of comprehensive, accurate and usable databases for sustainable materials design.

\end{abstract}


\noindent \textbf{Keywords}: Multicomponent alloys, High-entropy alloys, Materials databases, Large language models, Sustainable alloy design







\section{Introduction}
\input{}

Sustainable materials design must simultaneously satisfy two fundamental objectives: achieving target performance and reducing environmental, supply-chain, and socio-economic impacts. While decades of experimental and computational studies have generated a vast body of knowledge relevant to both objectives, especially for the former case, this information is largely dispersed across the literature in unstructured and heterogeneous format. As a result, it remains difficult to directly exploit this knowledge for systematic sustainable materials design. Converting unstructured literature knowledge into structured, machine-readable representations is therefore a critical first step toward enabling data-driven materials design strategies, including surrogate machine learning (ML) model based materials screening and model-guided experimental or computational workflows that could jointly and efficiently co-optimize performance and sustainability
\cite{saal2020machine,ramprasad2017machine,talluri2025discovery,cheetham2024artificial,handoko2025artificialintelligencegenerativemodels,nematov2025machinelearningdriven,srinivasan2020machine,balasubramanian2025machine,parambil2025transchem,kim2019active,pilania2016machine,pilania2013accelerating}. An inherent challenge with rapidly expanding materials science literature is that most of the measurements and results are reported in unstructured formats across text, tables, and figures \cite{katzer2025towards}. This makes manual organization of existing knowledge extremely tedious, time-consuming, and infeasible at scale \cite{jain2013commentary}, motivating the need for automated extraction pipelines that can transform literature data into structured, machine-readable databases \cite{smith2022challenges,hira2024reconstructing}. Coupling such automated pipelines with materials sustainability indicators can further enable accelerated discovery of materials that simultaneously meet performance requirements and reduced environmental impacts.

Early approaches on data extraction largely relied on task-specific ML models and pipeline-based or rule-driven methods for named entity recognition and relation extraction (NER/RE) \cite{zhu2021t, zhu2023evolution, yao2019docred, li2016biocreative}. While effective at identifying individual entities and predefined relations, such approaches can struggle to faithfully represent the structured, context-dependent relationships that are common in materials science and chemistry. In many cases, a reported property is meaningful only when interpreted jointly with a specific set of experimental conditions, where the relevant contextual variables may differ across properties and measurement protocols. For example, reporting that Al$_{0.1}$CrCoFeNi exhibits a corrosion potential ($E_{\text{corr}}$) of $-0.15$ V is insufficient without specifying the electrochemical environment, e.g., $E_{\text{corr}} = -0.15$ V at pH 2.43 in 0.6 M NaCl. In contrast, mechanical properties, such as yield strength, are typically meaningful only when accompanied with deformation conditions, including strain rate and temperature. Capturing such facts requires modeling relationships among multiple interdependent entities whose combinations and importance vary across studies. Although these dependencies can, in principle, be represented through compound or higher order relations, assembling adequate annotated training data that spans the diversity of properties, experimental contexts, and writing styles encountered in the literature remains a significant practical challenge. The advent of large language models (LLMs) has dramatically changed the landscape of the way in which the extraction task is handled because LLMs such as GPT-4, Claude 4, LLaMA-3, and Gemini 2.5 have demonstrated strong capabilities in leveraging semantic relationships across tokens in natural language sequences of varying length \cite{vaswani2017attention,comanici2025gemini, achiam2023gpt,grattafiori2024llama}. They have shown particular effectiveness in tasks such as text summarization, sentiment analysis, reasoning and problem solving \cite{guo2023evaluating, zaki2024mascqa, chowdhery2023palm, liu2022multi, hoffmann2022training, srivastava2023beyond}.

Given their success across a wide range of natural language tasks, there is growing interest in leveraging LLMs for automated scientific information extraction \cite{polak2024extracting, gupta2024data, gupta2024discomatdistantlysupervisedcomposition, dagdelen2024structured, duan2025llm, hira2024large, mishra2025foundationallargelanguagemodels, mandal2025autonomousmicroscopyexperimentslarge, choudhary2025chatgpt, schilling2025text, marini2025data, garcia2024review}. In an early effort, Dagdelen et al. explored fine-tuned GPT-3 for NER/RE to model complex entity relationships; however, the reported extraction accuracy and consistency were insufficient to support robust large-scale database construction \cite{dagdelen2024structured}. Subsequent work by Polak et al. \cite{polak2024extracting} demonstrated substantial improvements in extraction accuracy owing to advances in LLM capabilities, using sequential prompting workflows across both closed- and open-source models. While this study highlighted the promise of LLMs for materials information extraction, it reflects a broader trend in early efforts that emphasized pipeline development, rather than the construction of comprehensive databases intended for downstream usage. Efforts to scale extraction to larger datasets have been explored by Hira et al. \cite{hira2024large} and Gupta et al. \cite{gupta2024data}. Hira et al. trained graph attention networks to extract 18 properties from tabular data, while Gupta et al. fine-tuned GPT-3.5 with paragraph-level filtering to enable large-scale polymer data extraction across a broader fraction of the literature. Although these approaches produced substantially larger databases with improved accuracy, both were constrained to predefined and limited property sets (18 and 24 properties, respectively) and did not systematically capture materials-specific processing or testing conditions that are required to interpret measured properties, compare results across studies, and support downstream analysis and materials selection.

Notably, both Polak et al. \cite{polak2024extracting} and Gupta et al. \cite{gupta2024data} emphasized challenges related to data standardization, particularly the consistent representation of material compositions in the presence of aliases, acronyms, and heterogeneous nomenclature. Such issues complicate composition-based search, elemental chemistry analysis, and data reuse, and are an important consideration for the practical utility of automatically extracted materials databases. Finally, the cost of deploying LLM-based extraction workflows, whether through token-based APIs for proprietary models or GPU-intensive self-hosting of open-source models, remains a critical factor in determining their practicality for scalable database construction \cite{poddar2025towards, jiang2025demystifying, fernandez2025energy, samsi2023words}. Collectively, these studies highlight both the rapid progress enabled by LLMs and the need for cost-effective extraction workflows that produce standardized, compositionally precise, and context-rich datasets spanning a broad range of material properties.

Thus, the goal of this work is to develop a general-purpose extraction pipeline that is robust to diverse writing styles and reporting practices, captures a broad range of materials properties and experimental conditions, and produces standardized, compositionally precise databases suitable for sustainability-driven materials selection. We propose a workflow that leverages prompt engineering to enable accurate and fully automated data extraction using a conversational LLM. 
Specifically, the workflow employs in-context learning strategies, including system-level prompting, domain-specific technical context, few-shot examples, and chain-of-thought prompting \cite{dong2022survey, ouyang2022instructgpt,brown2020language, wei2022chain}. To mitigate hallucinations and missed entries, the prompts include explicit task instructions, technical definitions and commonly used terms, and manually curated annotations \cite{perkovic2024hallucinations}. Importantly, the prompt is tailored to each target paragraph using retrieval augmented generation (RAG) to dynamically select few-shot examples that are most similar to the target paragraph \cite{lewis2020retrieval}, thereby increasing data extraction efficiency as well as accuracy.

Overall, our data extraction workflow operates in two stages: (i) paragraph-level text extraction and (ii) table-based data extraction. In the first stage, alloy systems were identified along with reported properties and characterization techniques. Because such information is typically reported in abstracts and experimental sections, only these paragraphs were used for text extraction. Notably, numerical property values were not extracted at this stage. In the second stage, we extracted numerical property values of alloy systems from tables, where the majority of quantitative measurements are typically reported. Unlike prior studies, which focused on a narrow set of pre-defined properties, our workflow targeted over 350 distinct materials properties, while ensuring properties outside this list were assigned to an `Others' category. Importantly, this property list was curated from the database generated in the first stage, and was subsequently used to effectively guide table extraction in the second stage in a targeted manner. Each record in the second database includes alloy composition, processing conditions, testing conditions, property name, value, and unit, allowing several downstream tasks. The workflow resulted in two databases with 37,711 (first stage) and 148,069 (second stage) entries, derived from 10,829 literature articles. Further, the evaluation of our workflow against expert-annotated benchmarks, including two review articles, achieved at par or superior performance as compared to prior LLM works, with a precision and recall of 0.81 and 0.86 for text extraction, and 0.98 and 0.81 for table extraction, respectively.


Finally, we demonstrate the utility of such a comprehensive database by focusing on design of sustainable multi-component (or high-entropy) alloys in three distinct domains: lightweight structural alloys, soft magnetic materials, and corrosion-resistant alloys. The alloy properties obtained from the database were coupled with sustainability indicators (encompassing aspects of supply risk, environmental burden, and socio-economic factors) to find materials that are not only better from sustainability aspect but also display better property performance than current state-of-the-art commercial alloys in each of these domains. The unique attribute of our database to accurately capture alloy chemical composition enabled such a composition-resolved analysis and identification of potential sustainable alloys from existing literature. To facilitate reuse by the broader community, the curated database is made publicly accessible through an interactive web-based platform \href{https://alloy.tattvasar.com/}{(Alloy Tattvasar)}. 

We also highlight some of the key challenges in LLM-based automated data extraction from literature, especially in the context of alloys, including capturing property and composition data associated with multi-phase alloys, use of unconventional symbols or abbreviations, and quantitative information embedded in figures, among others. Overall, this work highlights the capability of LLM-based methods to extract complex, unstructured materials data and transform it into structured representations suitable for downstream AI/ML and materials discovery tasks. Beyond the case studies presented here, the database can serve as a foundation for learning composition-property relationships or training surrogate models to accelerate materials discovery. The framework is readily extensible to other material classes and can be coupled with rigorous sustainability methodologies, such as life cycle assessment, to support sustainable materials design.

\section{Results and Discussion}

\subsection{Overall workflow}

\begin{figure}[h]
    \centering
    \includegraphics[width=\linewidth]{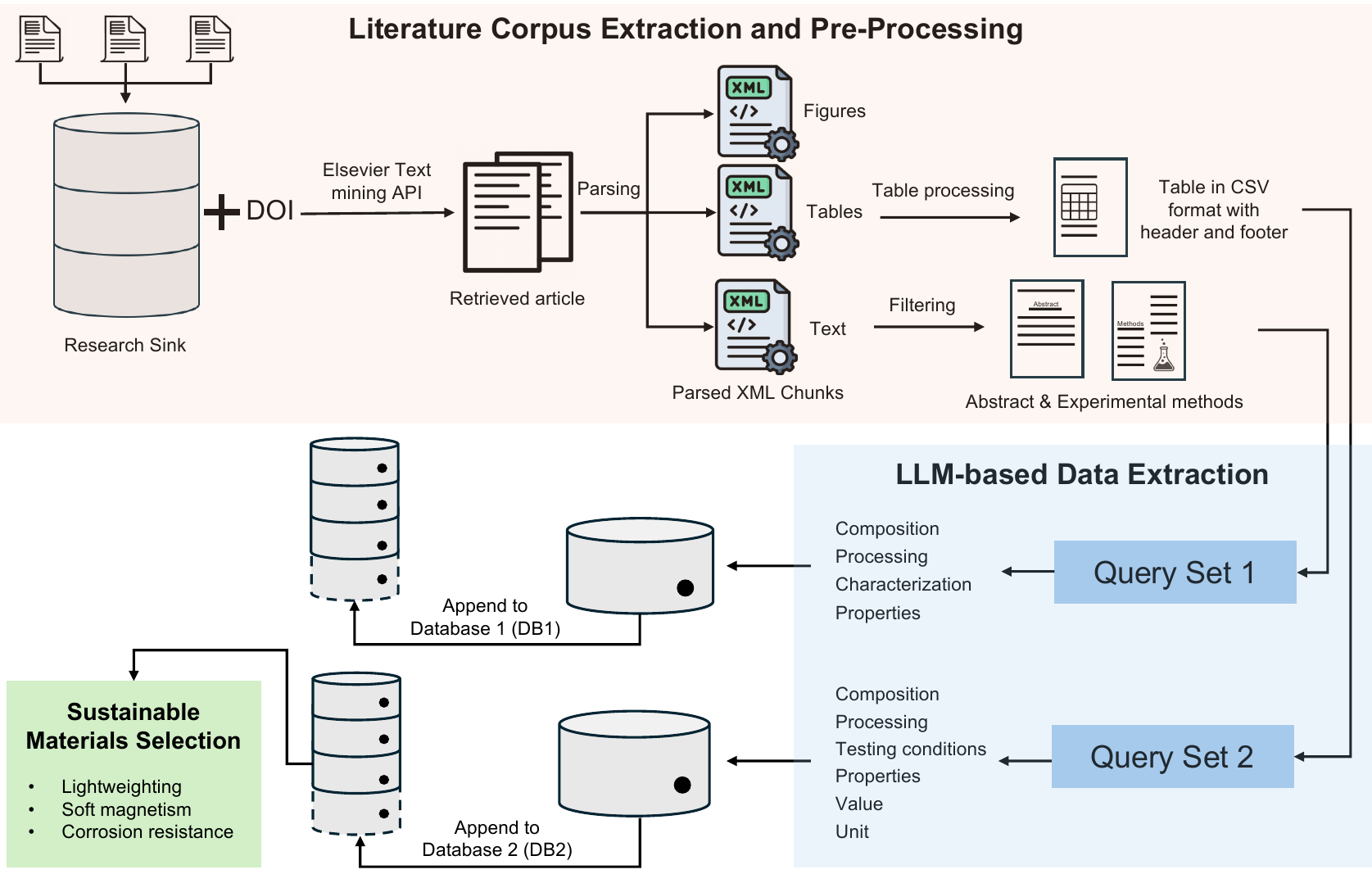} 
    \caption{
        Overall LLM-based alloy data extraction pipeline developed in this work. Starting from list of article DOIs, different sections of the article were separately extracted in XML format using publisher APIs. This data was selectively passed through two LLM extractor consisting of two query sets. While query set 1 (QS1) processed textual data from abstract and experimental section of the article, query set 2 (QS2) processed tabular data. Each of the query sets have their own series of prompts involving context, few-shot examples, etc. to finally output information such as alloy composition, properties, among others, in a tabular format. Mined data entries from each article were appended to associated databases, which were further cleaned to develop the alloy datasets. Finally, the QS2 database containing diverse property records of alloy systems was used for sustainable materials selection across 3 different application domains.
    }
    \label{fig: Fig1}
\end{figure}

Figure \ref{fig: Fig1} illustrates the overall workflow used for data extraction from research literature. A literature corpus was assembled by querying the Web of Science database with the term ``multi-component alloys” or ``high entropy alloy", which returned approximately 18,000 articles. Of these, 10,829 articles were accessible via Elsevier API which were were retrieved and converted from DOI to structured XML using an automated mining and parsing framework (see Methods section 4.1) \cite{jensen2019machine}. From this corpus, two complementary categories of information were extracted in form of separate databases
: (i) alloy system names and compositions, their characterization methods and reported property names, and (ii) quantitative property values and associated experimental conditions.

For the first database, a series of LLM prompts, termed query set 1 (QS1), were constructed to extract alloy systems, processing conditions, characterization techniques, and studied properties, without attempting to recover specific numerical values. Since such information is predominantly reported in abstracts and experimental sections, only these paragraphs were retained, filtered, and stored in CSV format before passing through the workflow. Each paragraph was analyzed using QS1, and the extracted outputs were post-processed to generate alloy-level records uniquely defined by composition and processing conditions for each article. Importantly, such database can be analyzed to understand which alloy systems have been studied, how they are characterized, and which properties are reported. Further, it can be used to collate a list commonly studied characterization techniques and reported property values, or to even navigate to relevant section of an article such as XRD spectra, optical micrographs, or other characterization images.


Similarly, for the second database, a complementary set of prompts, termed query set 2 (QS2), was developed to extract quantitative data. As numerical results are predominantly reported in tables, all tables from each article, along with their captions and footnotes, were retrieved and stored in a structured CSV format before passing through the LLM. The QS2 was specially tailored to extract reported property values and their units, testing conditions, and alloy composition and processing conditions. To improve extraction performance, the list of properties obtained from first database were used as part of QS2. Further, the outputs from QS2 were post-processed to have alloy composition and property values in standardized units, enabling detailed comparative analysis. Finally, the curated second database was coupled with composition-based sustainability indicators to identify potentially sustainable alloy candidates in various application domains.


\subsection{Query Set 1}
\subsubsection*{Query Set 1 Prompt Details}

\begin{figure}[h]
    \centering
    \includegraphics[width=\linewidth]{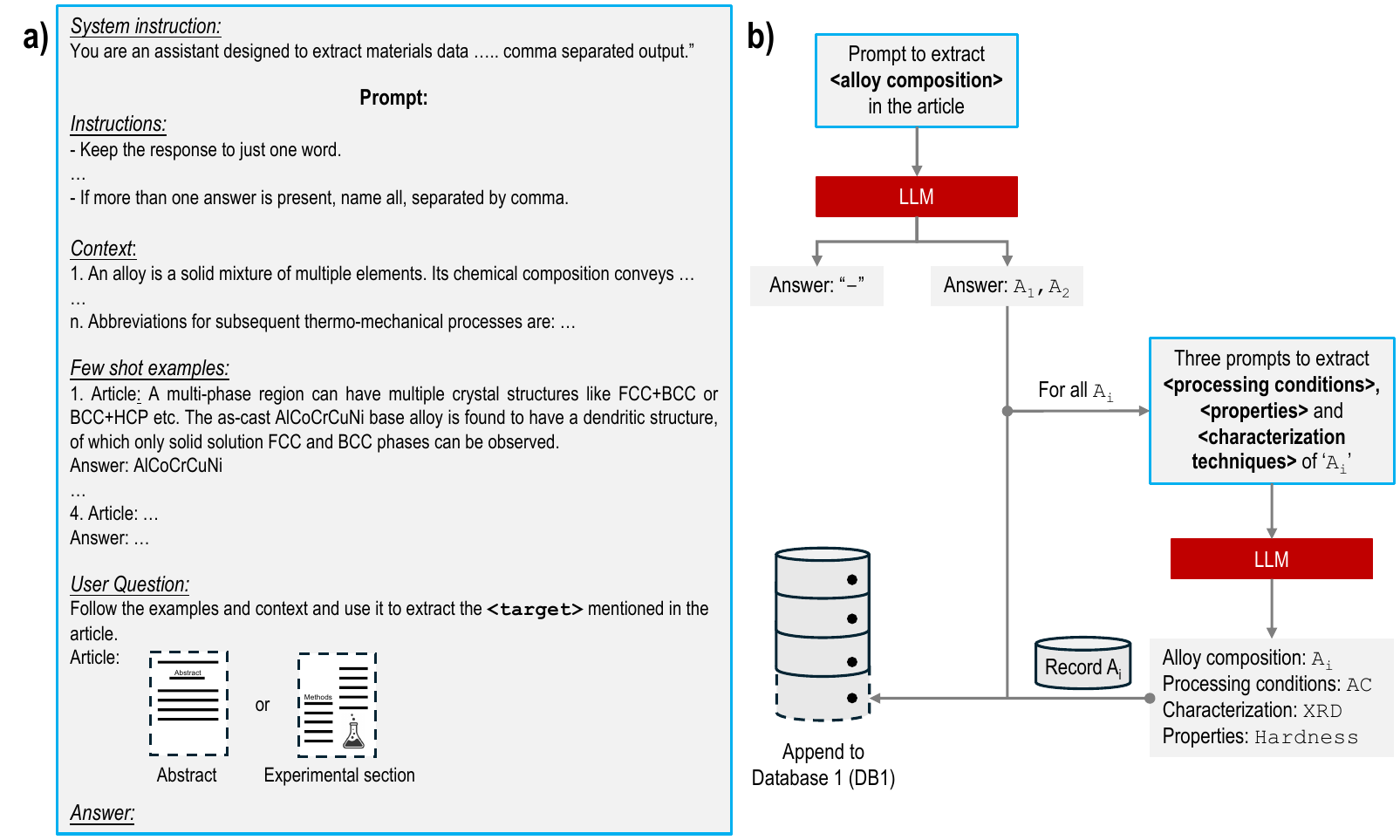} 
    \caption{
        a) Schematic of the QS1 prompt structure, comprising system and formatting instructions, domain-specific context, RAG-selected few-shot examples, and user query with the target article paragraph, organized in a chain-of-thought framework. b) QS1 data extraction workflow for a single article, in which alloy compositions are first extracted from each input paragraph, followed by extraction of the corresponding processing, characterization and property data, all of which are subsequently appended into DB1.
    }
    \label{fig:data_extraction_pipeline}
\end{figure}

Figure \ref{fig:data_extraction_pipeline}a details the prompt structure employed for QS1. The input to the LLM was composed of four components: system instructions, contextual definitions, few-shot examples, and formatting instructions, arranged in a step-wise, chain-of-thought style so that the model executes the task through discrete substeps. The system instructions explicitly define the LLM's role as a materials-data extraction expert as specifying such role and instruction tuning has been shown to substantially influence model behavior and alignment \cite{ouyang2022instructgpt}. The context block (termed as universal information in our prompt) provides concise domain definitions required to interpret common reporting conventions in alloy literature. For example, notation and interpretation of atomic ratios, typical format of reporting alloy compositions (e.g., fixed-stoichiometry expressions such as Al$_{0.3}$Cu$_{0.7}$ or parametric forms like Al$_x$Cu$_{1-x}$ with $x = 0.2, 0.3$), and common processing conditions such as vacuum-arc melting, hot isostatic pressing etc. The few-shot component contains a diverse pool of expert-annotated extractions from paragraphs in material science literature that demonstrate the desired extraction for different reporting styles. A total of 98 expert-annotated examples were constructed, embedded using OpenAI’s text-embedding-ada-002 encoder, and indexed into four separate vector databases corresponding to alloy composition, processing conditions, properties, and characterization techniques, with approximately 25 examples per category (see \href{https://github.com/MI-LAB-IITM/LLM_HEA_Data_Extraction/}{GitHub}). During extraction, RAG was employed to retrieve the top-5 most similar examples (by cosine similarity) to the user query, which were included as demonstrations in the prompt. This was done as empirical evidence indicates that few-shot examples and retrieval-augmented context substantially improve LLM performance on domain tasks \cite{brown2020language,lewis2020retrieval}.

To promote robust intermediate reasoning, the prompt was organized in a chain-of-thought or stepwise fashion, a technique known to elicit improved reasoning and extraction for complex tasks \cite{wei2022chain}. For QS1, the chain-of-thought reasoning unfolded in four stages: first, the model was guided to understand the extraction task and its objectives; second, it processed the background context and encouraged the model to become more attuned to materials literature and terminology; third, the annotations allowed the model to learn how an expert would perform the extraction; and lastly, it was presented with the target paragraph from which the data is to be extracted. Finally, strict formatting instructions were included in the prompt to obtain machine-parseable outputs and to simplify downstream post-processing. Specifically, these instructions included: (i) keeping the response to a single word; (ii) replying with ``\texttt{-}'' when context was insufficient or the answer was unknown; (iii) list multiple answers separated by commas when more than one applies; and (iv) strictly following the three subsequent steps embedded in the prompt. Together, these components produce structured paragraph-level outputs that were subsequently post-processed into unique alloy records (defined by composition and processing conditions) for the QS1 database. In regard to the type of LLM that was used, among the LLMs evaluated on smaller benchmark datasets (see Table S1), a combination of \texttt{GPT-4o} and \texttt{GPT-4o mini} provided an optimal balance between extraction accuracy and computational cost. This configuration was therefore selected for large-scale data extraction. More details on the different components of the prompt along with examples are included in Methods section 4.1.

Figure~\ref{fig:data_extraction_pipeline}b) illustrates the QS1-based extraction workflow. First, alloy compositions were identified across all paragraphs of each article. For each alloy, every paragraph in the article was examined to extract the associated processing conditions, characterization techniques, and reported properties. It should be noted that paragraphs may contain zero, one, or multiple alloy systems, which were handled through conditional skipping or iterative parsing. Because the same alloy system can be discussed across multiple paragraphs within an article, QS1-extraction often resulted in redundant entries. Thus, a post-processing step was conducted to capture unique information records using a combination of alloy composition and processing condition as a identifier. In cases where the processing condition was not explicitly mentioned, we assumed that the alloy was studied under a single processing route.

\subsubsection*{Query Set 1 Evaluation}
Next we wanted to validate the performance of QS1 on a manageable dataset before it can be used for large scale materials data extraction. However, there are no standard benchmark datasets that can be directly used for such a task (unlike other LLM domains of logical reasoning, general language understanding, etc.). Further, manually curating datasets also risk introducing biases toward certain types of articles, making the evaluation less reliable. Thus, a more balanced evaluation strategy was adopted using review articles, which are known to compile and organize data from a broad literature base. Specifically, for QS1, a review article \cite{miracle2017critical} was considered which covered information on alloy composition, processing conditions and properties, from around 121 original publications out of which 85 articles were available via the Elsevier API. These 85 articles were processed via the QS1 workflow, and the extracted outputs were compared against the consolidated tables reported in the review article. 



\begin{figure}[h]
    \centering
    \includegraphics[width=\linewidth]{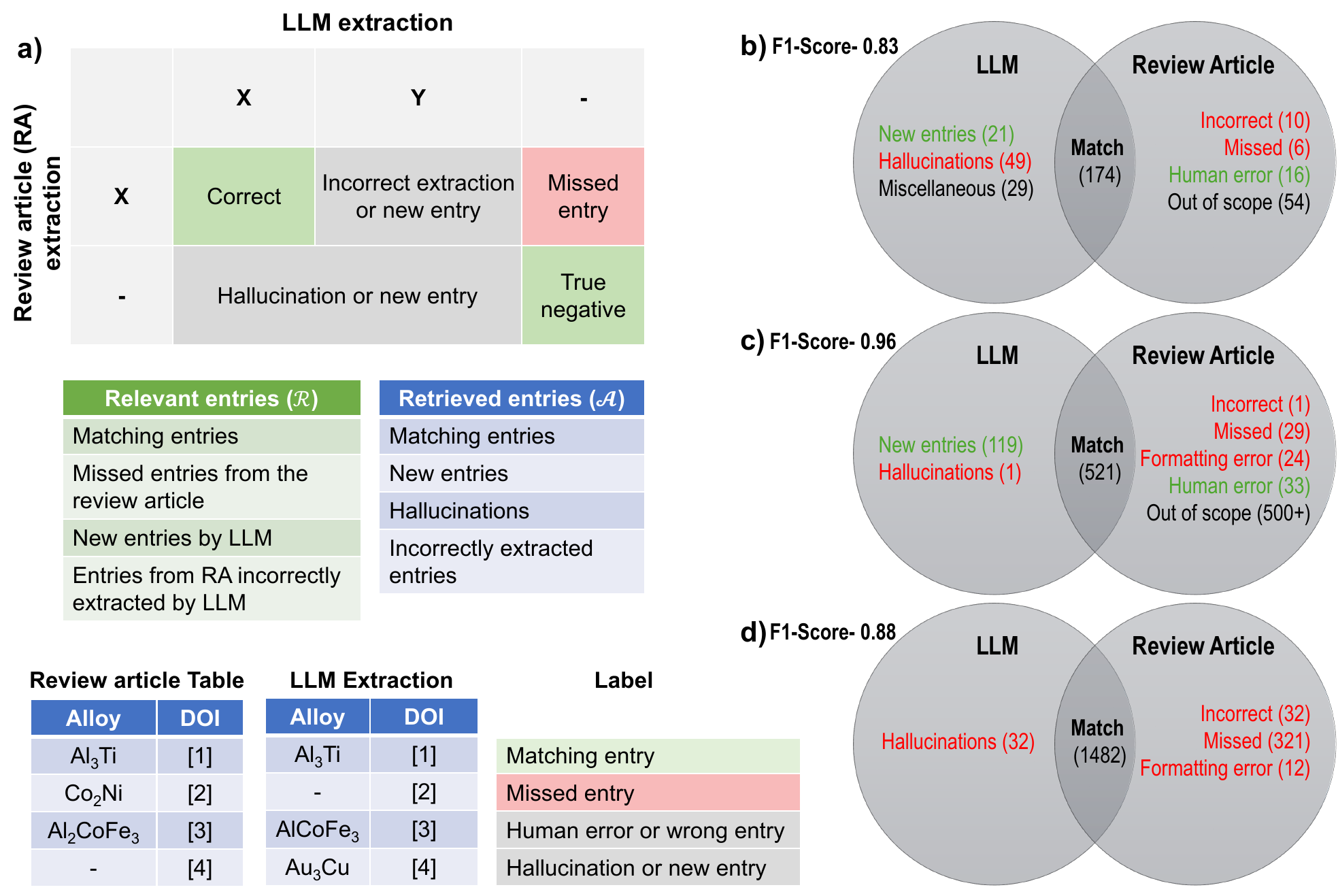} 
    \caption{
        a) Extended confusion matrix used in this work to evaluate QS1 and QS2. An example review article and corresponding LLM extracted datasets, capturing different types of relevant ($\mathcal{R}$)  and retrieved ($\mathcal{A}$) entries. Venn diagrams capturing evaluation of b) QS1 on review article dataset, and of QS2 on c) review article and d) manually curated datasets.  
    }
    \label{fig:modified_confusion_matrix}
\end{figure}

Typically, evaluation of a data extraction task is done using confusion matrix and associated error metrics such as precision, recall, and F1-score \cite{pearson1905general}. However, in our case, a standard confusion matrix was found to be insufficient because the assumed ground truth, i.e., the values reported in review article tables, were not always correct. Thus, discrepancies between QS1 extraction and review article values did not necessarily mean the LLM extraction was a hallucination or an incorrect extraction. It could also mean that the review article ground truth was incorrect. For example, as shown in Figure~\ref{fig:modified_confusion_matrix}a, the review article reports Al$_2$CoFe$_3$, whereas QS1 extracts AlCoFe$_3$. This mismatch may arise from either an incorrect/hallucinated extraction by the LLM or due to inaccuracy in the review article itself. Similarly, we also observed cases, termed new entries, where QS1 identified valid records absent from the expert-curated dataset; see example extraction of Au$_3$Cu in Figure~\ref{fig:modified_confusion_matrix}a. To account for all such scenarios an extended confusion matrix was considered as shown in Figure \ref{fig:modified_confusion_matrix}a, which explicitly captures mismatches, omissions, and novel extractions. 
Performance metrics were computed as $\mathrm{precision} = \frac{|\mathcal{R}\cap\mathcal{A}|}{|\mathcal{A}|}$, $\mathrm{recall}= \frac{|\mathcal{R}\cap\mathcal{A}|}{|\mathcal{R}|}$ and F1-score $=\frac{2\,|\mathcal{R}\cap\mathcal{A}|}{|\mathcal{R}|+|\mathcal{A}|}$, where $\mathcal{R}$ is the set of relevant (or correct) records in the review article as well as those records that are missing in the review article, but correctly extracted by the LLM, and $\mathcal{A}$ is the set of retrieved records extracted by the LLM, as captured in Figure \ref{fig:modified_confusion_matrix}a.

Overall, the QS1 workflow achieved an F1-Score of 0.83 on the review article, with details on the different kinds of extraction errors shown using the Venn diagram in Figure \ref{fig:modified_confusion_matrix}b. 
Relatively large number of hallucinations were observed in the QS1 outputs. 
These hallucinations were primarily due to the difficulty LLMs faced in performing multi-step arithmetic and non-trivial symbolic reasoning. For instance, consider the input: ``We synthesize the aluminum alloys, Al$_x$(CoCrCuFeMnNiTiV)$_{100-x}$ (x = 11.1, 20, 80 at\%) and processed them by annealing at various temperatures and cold-rolled the alloys to varying degrees as indicated in Table~3.'' Here, the model must first recognize that multiple alloys are described, substitute the values of $x$, distribute the remaining $100-x$ equally among eight elements (Co, Cr, Cu, Fe, Mn, Ni, Ti, V), and then present the resulting compositions in proper chemical notation. Despite exposure to similar examples in few-shot prompting, the absence of explicit step-by-step demonstrations made such reasoning particularly challenging, leading to inaccurate or hallucinated outputs.

At the same time, there were new entries that QS1 extracted but were missing in the expert-compiled review article. For example, when given the input: ``A new high-entropy alloy system was designed based on AlCoCrCuFeNi, by replacing Cu with Mo to improve strength and thermal stability,'' the LLM correctly extracted both AlCoCrCuFeNi and AlCoCrMoFeNi, although the latter was omitted from the review article \cite{hsu2010effect}. However, as substitutions grow more complex, extraction accuracy declines. In the sentence: ``To further understand element influence on alloy property, in CoCrFeMnNi alloy system we synthesize a new system by substituting elements as Ti for Co, Mo or V for Cr, V for Fe, and Cu for Ni for wear-resistance study,'' the model produced erroneous compositions (e.g., CoCrMnMoNi) due to difficulties in handling multiple substitutions simultaneously \cite{otto2013relative}. Finally, we also encountered instances where the LLM outperformed the review articles in extracting detailed processing conditions. While some conditions were incorrectly curated in the review article tables, the LLM was able to identify and record them correctly. Since the LLM is provided only with paragraphs from the abstract and experimental methods, some entries reported in the review article cannot be inferred from this information alone. These entries are therefore classified as out of scope and were excluded from the error analysis. These examples highlight both the limitations of the current LLMs in handling multi-step compositional logic and their potential to complement expert curation by capturing valid entries overlooked in manual reviews.

 
\subsubsection*{Query Set 1 Database}

With the QS1 pipeline validated, it was applied to the full corpus of 10,829 multi-component and high-entropy alloy articles to construct a comprehensive database, termed database 1 (DB1). In total, 37,711 alloy system records were extracted, of which 36,536 (97\%) were deemed directly usable. Here, direct usability denotes alloy compositions expressed in standard chemical notation, comprising valid elemental symbols with optional subscripts denoting atomic ratios. Such representations are critical for downstream materials discovery and computational analysis, as they enable reliable parsing and composition-based screening. For example, Al$_3$Ti is directly usable, whereas extractions such as ``Al-HEA” or ``Al-based alloy” are not. Usable compositions were identified using a rule-based pattern matching script.

While more details on the extracted DB1 are provided in SI, here we highlight some of the important trends. The DB1 contains 15,998 unique alloy compositions. 
Nickel, iron, and chromium were found to be the three most common alloying elements, while carbides, borides, and oxides frequently occurred as additives. On average, each article contained 3.02 unique alloy compositions, which is partly inflated due to inclusion of related alloys from other refereed works. Although multi-element alloys often contain more than three components, certain element pairs such as Fe-Ni, Co-Ni, and Cr-Ni were found to occur together frequently. Other commonly co-occurring elements are also highlighted in Figure S1c. A word cloud plot was generated to visualize the most frequently studied properties (Figure S1d) and characterization techniques (Figure S1e). As expected, mechanical properties such as hardness, ductility, and strength were found to dominate. Similarly, X-ray diffraction, scanning electron microscopy, and transmission electron microscopy were found to be the top characterization methods.


\subsection{Query Set 2}
\subsubsection*{Query Set 2 Prompt Details}


For QS2 we focused on extracting quantitative information from tables. Unlike prior works, we did not want to limit our extraction to a narrow set of pre-defined properties but wanted to develop a generic pipeline. However, there are inherent challenges associated with extracting property values from tables. For instance, property names are often reported in an inconsistent manner (e.g., ``Vickers hardness'', ``Vickers-hardness'', ``microhardness'', ``micro-hardness'', ``hardness'' all refer to the same property). Further, properties may be represented using symbols or abbreviations whose definitions are provided elsewhere in the manuscript or inferred from context. Thus, to enable reliable extraction, we first curated a controlled vocabulary of 354 materials properties derived from DB1 using a combination of LLM-assisted clustering, algorithmic filtering, and manual validation. DB1 initially yielded over 8,000 unique property descriptors, many of which were semantically redundant (e.g., ``ductility,'' ``strain at failure,'' ``elongation before failure,'' and ``failure strain''). Simplifying this redundancy into a non-overlapping property set was essential to make the QS2 extraction consistent and exhaustive (discussed later). Nevertheless, not all DB1 properties could be mapped to this vocabulary. For example, properties like ``Epp–Eprot'' commonly reported in corrosion studies, or article-specific measures such as inclusion or bubble sizes, are not standardized across the field. To accommodate such cases, an additional ``Others'' category was included to capture properties outside the curated list. The Others category also helps mitigate ambiguities when the LLM cannot confidently classify a property, ensuring that potentially useful but non-standard or unclear entries were still retained rather than discarded.

\begin{figure}[h]
    \centering
    \includegraphics[width=\linewidth]{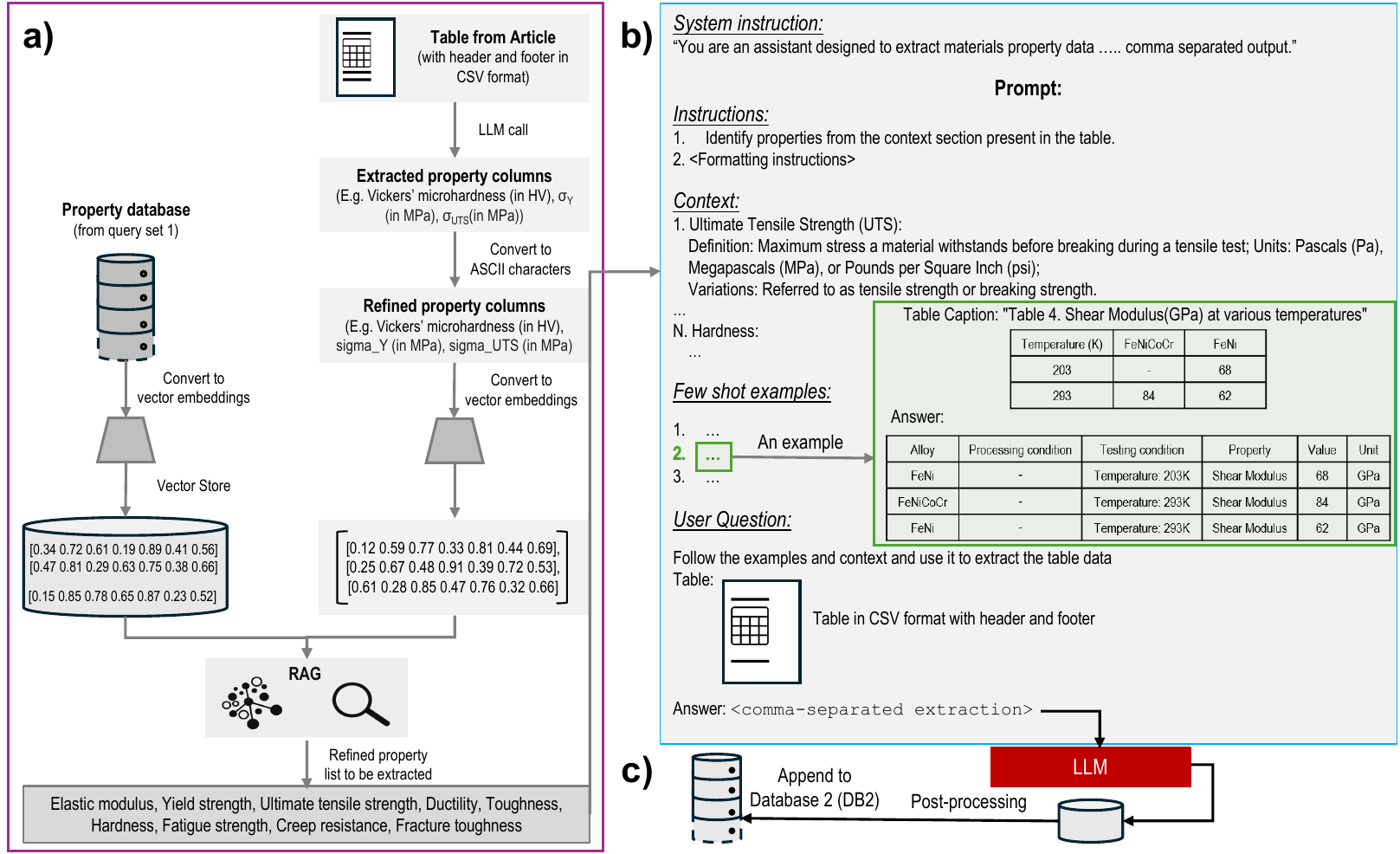}
    \caption{
        Overall framework for QS2. a) First, a narrowed list of property set were identified in the target article table using a modified RAG approach and a master list of 354 materials properties constructed using DB1. b) This selected property list was included as context in the QS2 prompt, which also included other aspects of prompt engineering such as chain-of-thought prompting, RAG-based few-shot examples and formatting instructions. An example few-shot example for QS2 is also included. c) Data records output by QS2 were post-processed for any formatting errors before being appended to DB2. See main text for more details on QS2 workflow. 
    }
    \label{fig:table_data_extraction_pipeline}
\end{figure}

Figure~\ref{fig:table_data_extraction_pipeline} illustrates QS2 pipeline for extracting information from tables. The key idea is to have two sequential LLM calls, first to extract standardized property names present in the table (while accounting for different symbols and property names) and then a second LLM call to extract information associated with the identified properties. For this, first all tables were downloaded article-wise in a CSV format. Each table was then processed by an advanced LLM to return cells and caption strings that may contain material properties. A detailed system prompt was employed for this step (see Methods section 4.2). As stated earlier, one major challenge was that a single property can be reported in many different ways; for example, the ultimate tensile strength can be written as $\sigma_\mathrm{UTS}$, $\sigma_\mathrm{max}$, or by its full name. Thus, a direct mapping of extracted cells to the master 354 property list was unreliable; even embedding cosine similarities showed poor performance because of presence of symbols and drastically different tokens for the same property.

To address this, a modified RAG approach was developed. The master list of 354 properties was first expanded to include alternative names and commonly used symbols, all normalized to ASCII characters (e.g., $\sigma_\mathrm{y}$ was converted to \texttt{sigma\_y}); see SI section 4 for details. Each property-notation pair was then embedded and stored into a vector database, which contained thousands of such representations. When the LLM returned candidate table cells, a similarity search was performed against this database to identify the top matches, as shown in Figure \ref{fig:table_data_extraction_pipeline}a. For each cell, the top three most similar property embeddings were retrieved. These matches were consolidated across the table to generate a reduced candidate set of approximately 10–20 properties, depending on table size and number of duplicates. This narrowed property list, along with concise property descriptions including definitions, typical measurement methods, testing conditions, and units, was then provided as context for the second LLM call to extract quantitative values. 
For the first LLM call, GPT-4o was used for all the tables. For the second LLM call, tables containing fewer than 30 cells were processed using GPT-4o~mini, while larger tables were handled using GPT-4o. This is mainly because of the comparable accuracy but lower cost of GPT-4o~mini observed in preliminary evaluations. 

The prompting strategy for QS2 followed the same principles as QS1, with a structured prompt containing few-shot examples, explicit formatting instructions, and chain-of-thought prompting; see Methods section 4.1 for details. Formatting instructions ensured that the LLM outputs were in a CSV format consistently. To further improve reliability, the LLM was instructed to match extracted properties against the narrowed set of candidate names. In cases where the retrieved set was incomplete (e.g., when Tafel slope was clearly present in the table but absent from the retrieved candidates), the property was labeled under the ``Others'' category. This approach allowed us to capture both standard and non-standard property names without discarding useful information. A sample few-shot example is shown in Figure~\ref{fig:table_data_extraction_pipeline}c, which contains a table reporting shear modulus values of two alloy systems measured at different temperatures. The expert-annotated example demonstrates that shear modulus must be extracted along with the corresponding testing conditions. 
Similarly, we also extracted processing conditions reported alongside alloy compositions. 
Capturing such information helps to build a more comprehensive alloy database. 


\subsubsection*{Query Set 2 Evaluation}

Similar to the case of QS1, QS2 was also evaluated on a review article \cite{george2020high} containing mechanical properties of high entropy alloys with around 1500 records. Additionally, a test set of $\sim$1800 target records across 33 research articles was manually curated.
The same evaluation metric used for QS~1 was applied to assess table extraction performance.  For mechanical properties, the pipeline performed particularly well, achieving an F1-score of 0.96, as shown in Figure~\ref{fig:modified_confusion_matrix}c, which is encouraging given that the mechanical properties dominate the HEA literature. When evaluated on a broader test set spanning mechanical, corrosion, electrical, optical, and other properties, the pipeline achieved an F1-score of 0.88 (Figure~\ref{fig:modified_confusion_matrix}d), supporting its suitability for large-scale literature mining. These scores are comparable to, and in several cases exceed, reported performances of related ML-based extraction approaches, which typically range from 0.6 to 0.9 \cite{polak2024extracting, gupta2024data, gupta2024discomatdistantlysupervisedcomposition, dagdelen2024structured, hira2024large}. However, a direct comparison of errors across different works should be interpreted with caution, owing to differences in scope, dataset composition, and the evaluation strategies.

An important observation regarding the performance of the QS2 is that despite its near-perfect precision, the pipeline missed approximately 20\% of the target records, leading to reduced recall. This behavior arises primarily from two factors: (i) the complex and varied nature of property descriptions in tabulated data, and (ii) the prompt instruction directing the LLM to omit entries for which it is uncertain. Certain properties are reported using non-standard or highly specialized nomenclature that may not be present in the vector store used for retrieval, resulting in incomplete identification of the expected 10-20 property names per table. Examples include rarely reported quantities such as $E_{\mathrm{sp}}$, the secondary passivation potential, and $\mu_{\mathrm{ss}}$, the steady-state coefficient of friction, which consequently have very few entries in the database.

Some other issues with QS2 outputs were also encountered. Despite including strict formatting instructions to generate comma-separated six output fields, 
around 2\% of the LLM produced outputs that could not be directly ingested into database. In some cases, missing contextual information, such as unavailable processing or testing conditions, resulted in misaligned columns. Although the model was instructed to insert a placeholder ``-'' in such cases, this was not always followed, resulting in incomplete records (e.g., \texttt{AlCoNi\(_{0.5}\)Ti, as-cast at 900K, Hardness, 900, MPa}). Additionally, descriptive processing or testing conditions themselves contained commas, further disrupting the expected format. While these inconsistencies prevented automated ingestion, post-processing scripts were used to reformat and recover relevant information. 
Another source of error arose from alloy composition extraction. Some studies reported compositions in weight percent rather than atomic percent, while others specified minor constituents using prefixed numerical notation (e.g., \texttt{Ti-6Al-4V}). Such cases contributed to both incorrect extractions as well as a corresponding LLM hallucinated record in Venn diagram analysis. Similar to QS1, the review article 2 used to evaluate QS2 responses too had out of scope entries. 
These correspond to records that could not be extracted using tabular information alone, as part or all of the required data appeared in the main text or figures. Thus, future workflow on development of such databases should incorporate strategies to systematically handle information distributed across different sections of an article. 


\subsubsection*{Query Set 2 Database}
Following validation, all tables from the 10,829 journal articles were considered for QS2. However, not all tables were found to be usable as some were available as images rather than structured XML. Ultimately, a total of 22,397 individual tables from approximately 6,000 articles were processed using QS2, out of which 14,001 tables yielded non null outputs. The pipeline extracted 148,069 records spanning a wide range of properties measured under diverse testing conditions across numerous HEA systems. However, many extracted alloy records used non-standard or article-specific nomenclature (such as aliases, commercial names), with only $\sim$78,000 records being directly usable. Therefore, a multi-stage post-processing pipeline (see Methods section 4.3) combining rule-based resolution, physical validation, and context-aware LLM refinement was applied to standardize compositions, increasing the number of usable alloy entries to 102,134 entries and substantially improving database completeness. The resulting database, termed database 2 (DB2), enabled large-scale analysis and visualization of property trends, including materials selection plots, and formed the basis for subsequent sustainability case studies presented in the following section.

\begin{figure}[h]
    \centering
    \includegraphics[width=\linewidth]{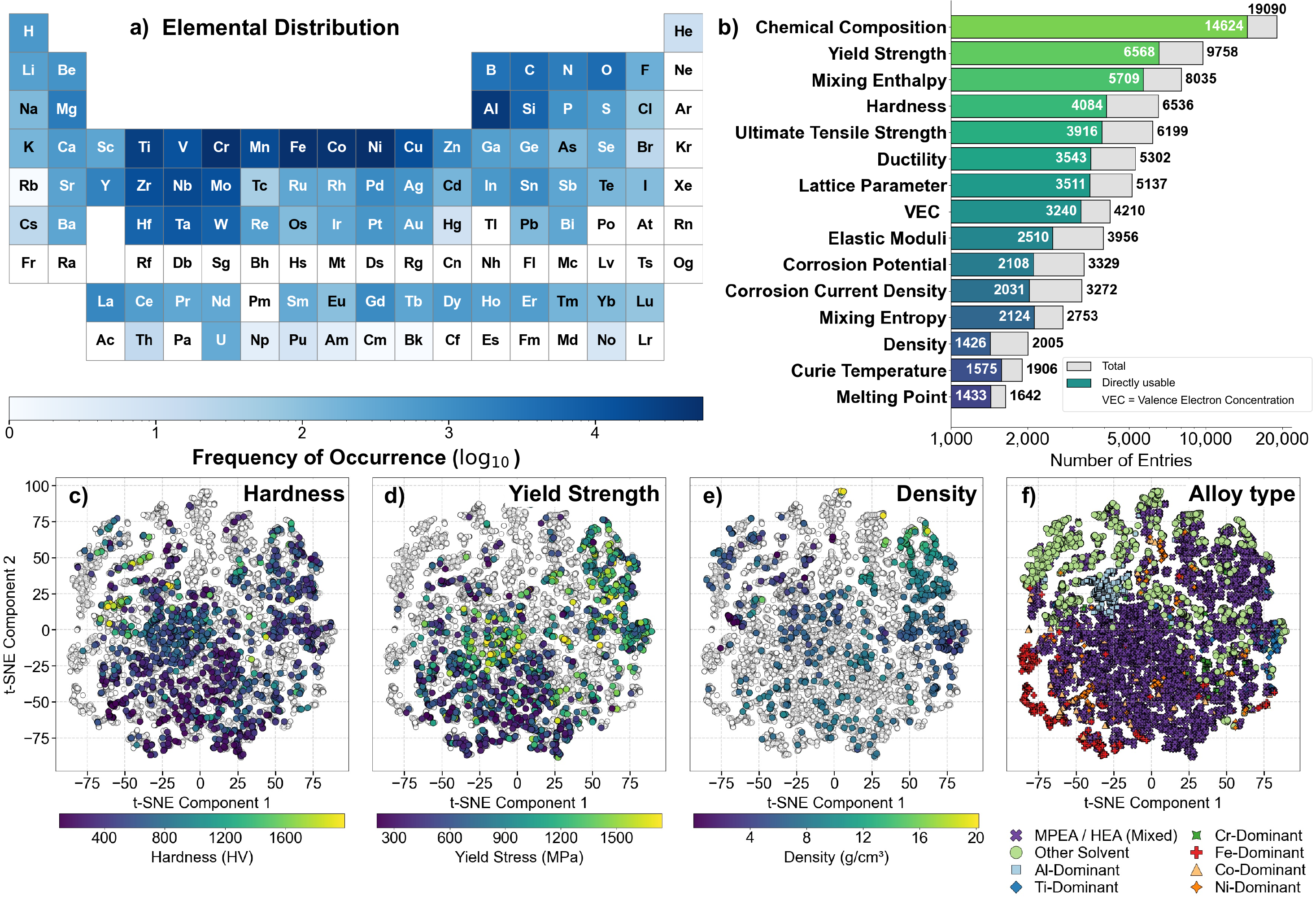} 
    \caption{Visualization of the LLM-mined DB2.
    a) Periodic-table heatmap showing the frequency of occurrence of elements in the database.
    b) Log-scale distribution of the 20 most frequently extracted material properties.
    t-SNE visualization of the full database, with colors denoting c) hardness, d) yield strength, e) density and f) family of the alloy.
    }
    \label{fig:Q2_metadata}
\end{figure}

The DB2 comprised of 16,381 unique alloy compositions. Among the standardized entries in DB2, the frequency of occurrence of each element in the periodic table was quantified, as shown in Figure~\ref{fig:Q2_metadata}a. Transition metals such as Ti, Cr, Fe, Co, and Ni were the most prevalent, with Al also appearing frequently, consistent with the trends observed in DB1. Figure~\ref{fig:Q2_metadata}b shows the most frequently reported properties in the database. As expected, mechanical properties, such as yield strength, hardness, and ultimate tensile strength, were the most prevalent. Corrosion-related metrics such as corrosion current density and corrosion potential and intrinsic properties including melting point and Curie temperature were frequently occurring too. These properties will be used later for sustainable materials design. Experimental compositional measurements were also frequently extracted as a part of DB2. 

In order to better visualize the database, elemental Magpie features \cite{ward2016general} were computed for each alloy by weighting individual elemental features by their atomic fractions. The resulting feature vectors were embedded into two dimensions using t-SNE~\cite{maaten2008visualizing} as shown in Figure~\ref{fig:Q2_metadata}c-f. To examine property distributions within this reduced space, three representative properties, i.e., hardness, density, and yield strength, were used to color-code alloys for which data was available. Distinct regions of low property values and continuous gradients corresponding to increasing property magnitudes can be observed. The availability of exact compositions further enables composition-property analyses in addition to property-property correlations, as discussed in Figure S2. Alloys were additionally classified into compositional families based on their dominant element, defined as an atomic fraction exceeding 35\% for alloys containing more than three elements. The resulting alloy family and property maps reveal interesting trends. Alloys from different families can be seen to cluster into different regions (Figure \ref{fig:Q2_metadata}f), with HEA occupying the central region of the plot. Further, as expected, regions of low density can be observed to coincide with Al- and Ti-based alloys. Similarly, low hardness and low yield strength regions correspond to Al-dominant regions. Property-property correlations are also evident from these visualizations. In particular, directions of increasing hardness tend to coincide with directions of increasing yield strength in the reduced space, indicating a correlation between these properties. 
Additional correlations revealed by these embeddings are presented in Figure S2.



\subsection{Sustainable Alloy Design}


We next illustrate the utility of the mined database for sustainability-driven materials selection through three representative case studies: lightweight structural alloys, soft magnetic materials, and corrosion-resistant systems.  To quantify the sustainability of a HEA, we integrate a set of nine curated sustainability indicators \cite{gorsse2025sustainability} covering eighteen alloying elements most frequently reported in HEA research. These indicators capture supply risk, environmental burden, and socio-economic considerations related to element production, including the reliance on strategic or critical raw materials. By normalizing and aggregating these metrics at the alloy level, we derive a sustainability index that enables direct comparison across different alloy compositions. Further, in each case, benchmark alloys that represent some of the top performing alloys from a functional standpoint are also included to highlight design of alloys that not only perform better from a sustainability standpoint but also from functional performance viewpoint. It should be noted that both lightweighting and corrosion resistance (proxy to durability) are themselves key sustainability considerations as they affect both material usage and lifecycle performance.

\begin{figure}[h]
    \centering
    \includegraphics[width=\linewidth]{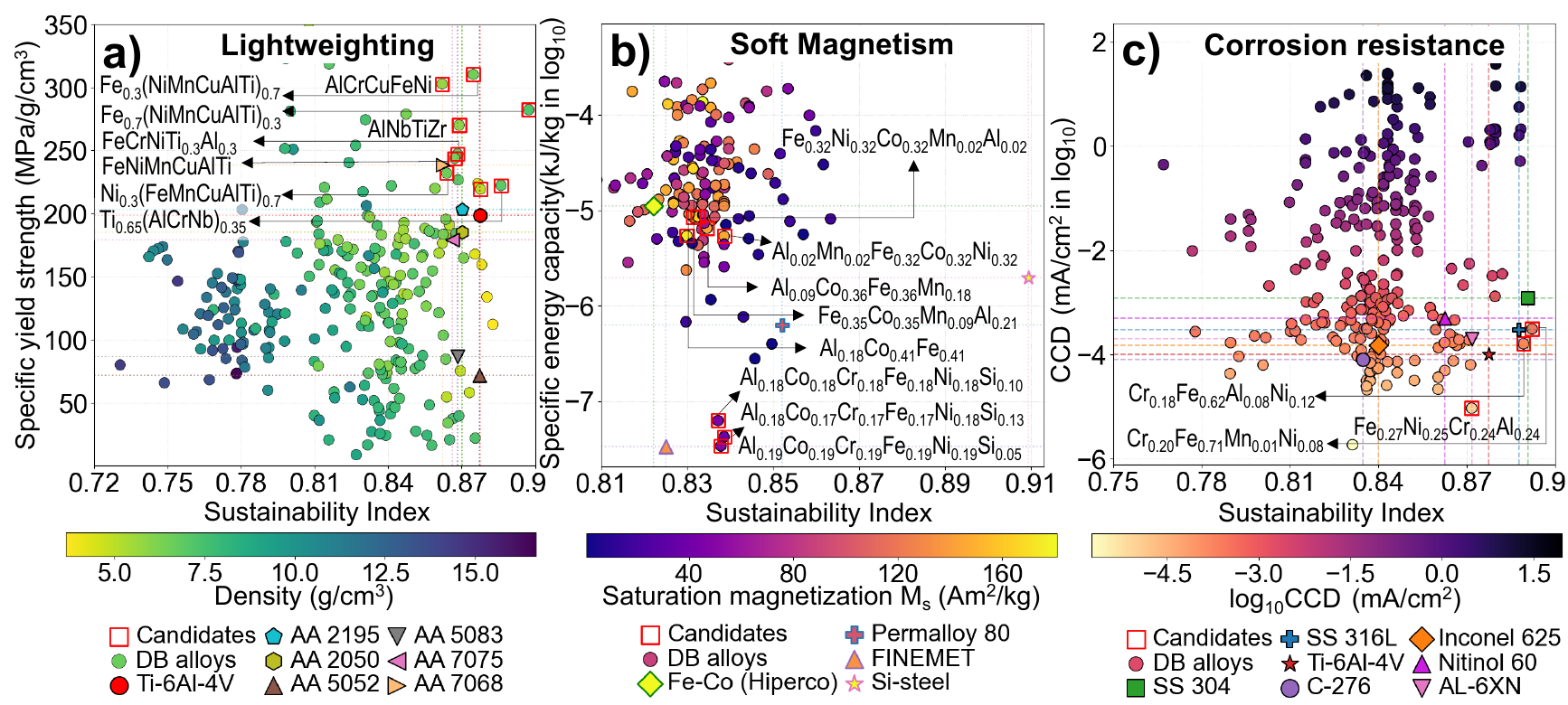} 
    \caption{
        Sustainability maps highlighting alloy designs satisfying performance and sustainability criteria across three application domains a) lightweighting, b) soft magnetism, and c) corrosion resistance. Promising candidate alloys are indicated by red square markers. CCD: corrosion current density.
    }
    
    \label{fig:Case_Studies}
\end{figure}

\subsubsection*{Lightweight functional materials}
As a first case study, we consider sustainability-informed alloy design in the context of lightweighting, motivated by its critical importance across applications such as automotive body structures, aerospace airframes and fuel tanks, marine vessels, and energy infrastructure. Across these domains, specific strength, defined here as the ratio of yield strength to density, is a widely accepted performance metric for materials selection, as it simultaneously reflects load-bearing capability and mass efficiency (related to fuel and handling costs). From DB2, we identified 262 alloy systems for which both yield strength and density were available and for which sustainability indices can be computed, enabling construction of a materials selection chart as shown in Figure \ref{fig:Case_Studies}a. For reference, several benchmark alloys spanning diverse lightweighting contexts, including aerospace-grade AA2195 \cite{kim2016effect} and AA2050 aluminum alloys \cite{lequeu2010aluminum}, automotive and marine environments related corrosion-resistant and highly formable AA5052 \cite{aziz2022mechanical} and AA5083 aluminum alloys \cite{jebaraj2020mechanical}, high-strength AA7075 \cite{jha2008metallurgical} and AA7068 aluminum alloys \cite{sathish2021investigation}, and the widely used titanium alloy Ti–6Al–4V \cite{boyer1996overview} are also included.
Notably, nine HEA compositions (marked red) emerged as promising candidates that not only exhibited competitive specific strength but are also composed of elements associated with favorable sustainability indices. These alloys were predominantly Al- and Ti-rich lightweight HEAs, consistent with their high specific strength and the low criticality, high abundance in Earth's crust, and established recyclability of Al and Ti, which collectively contribute to elevated sustainability scores. Notably, several high-performing candidates fall within the Ni–Mn–Cu–Al–Ti–Fe compositional space; for example, $\mathrm{Fe}_{0.7}(\mathrm{NiMnCuAlTi})_{0.3}$ exhibited competitive specific strength alongside favorable sustainability indices, driven by reduced fractions of more critical elements such as Ni and Cu  \cite{kondapalli2023elemental}. An additional sixteen compositions displayed substantially higher specific strengths but were not included in the present comparison due to the absence of sustainability indicators for one or more constituent elements.  As a consistency check, well-established lightweight HEAs such as AlCrCuFeNi, reported for electrical transmission applications \cite{chen2024synergistic}, and AlNbTiZr, known for its high specific strength and thermal stability \cite{jayaraj2018microstructure}, were recovered within this analysis; see Table S4 for the complete list of identified promising candidates. Beyond identifying promising candidates, the database enables traceability to original synthesis routes, processing conditions, and testing environments reported in the literature, while the diversity of source studies ensures that underrepresented alloy families are captured in the selection space.

\subsubsection*{Soft magnetic materials}

We next examine the design of soft magnetic materials, which are critical for applications such as electric motors, transformers, inductors, magnetic shielding, and power electronics, where efficient magnetic response under alternating magnetic fields is required. Desirable soft magnetic performance requires high saturation magnetization with low specific energy capacity, a measure of hysteretic loss that can be approximated by a scaled product of saturation magnetization and coercivity. 
Using DB2, we identifed 143 alloy compositions for which saturation magnetization, coercivity, and sustainability indices were available, enabling construction of sustainability map for soft magnetic performance as shown in Figure \ref{fig:Case_Studies}b. For industrial reference, we included several benchmark material classes, including Fe-Co alloys for high-saturation and high-temperature applications \cite{willard1998structure}, Fe-Si steels for cost-effective electrical machinery \cite{ouyang2019review}, Permalloy 80 (Fe-Ni) for high initial permeability and low coercivity in magnetic shielding and sensing \cite{cullity2011introduction}, and FINEMET-type nanocrystalline alloys for low-loss, high-frequency power electronics \cite{herzer1990grain}. Collectively, these benchmarks capture the diversity of operating conditions, frequency regimes, and processing constraints encountered in soft magnetic applications. These materials also illustrate the application-dependent trade-offs inherent to soft magnet selection: nanocrystalline alloys such as FINEMET offer exceptionally low hysteretic losses and superior high-frequency performance but require complex and costly processing routes; Fe-Co alloys deliver very high saturation magnetization and excellent thermal stability at the expense of increased hysteretic losses and reliance on Co, which is expensive; Fe-Si steels remain widely deployed due to their low cost, ease of processing, and scalability despite being optimized for moderate operating conditions; and Permalloy-type Fe-Ni alloys provide a balanced combination of low coercivity, manufacturability, and comparatively improved sustainability relative to nanocrystalline systems. A total of 11 promising candidate alloys (indicated by red square markers in Figure \ref{fig:Case_Studies}b) with performance comparable to the soft magnetic benchmarks were identified using DB2. Several Al-Co-Cr-Fe-Ni-Si high-entropy alloys (see Table S3) occupy the performance-sustainability space between Permalloy 80 and FINEMET, suggesting potential pathways to bridge the gap between low-loss performance and more sustainable production. In particular, the $\mathrm{Al}_{0.19}\mathrm{Co}_{0.19}\mathrm{Cr}_{0.19}\mathrm{Fe}_{0.19}\mathrm{Ni}_{0.19}\mathrm{Si}_{0.05}$
alloy showed comparable performance to FINEMET, while having a more favorable sustainability index owing to avoiding Nb usage. Among compositions clustered near the Fe–Co benchmark, Al$_{0.18}$Co$_{0.41}$Fe$_{0.41}$ emerged as a promising candidate for high-temperature soft-magnetic applications. It exhibits comparable saturation magnetization to Fe$_{0.5}$Co$_{0.5}$, and a lower specific energy capacity relative to Fe–Co alloys, while achieving an improved sustainability index due to reduced Co content. In addition, seven other candidates displayed favorable magnetic performance but were not included in the plot due to the absence of sustainability indices for one or more constituent elements. Beyond intrinsic magnetic properties, operating conditions impose additional constraints, particularly Curie temperature for high-temperature applications. Al$_{0.18}$Co$_{0.41}$Fe$_{0.41}$ exhibits a Curie temperature of 1254~K, which is nearly the same as Fe$_{0.5}$Co$_{0.5}$ \cite{wang2024developing}. Similarly, Fe$_{0.32}$Ni$_{0.32}$Co$_{0.32}$Mn$_{0.02}$Al$_{0.02}$ and Al$_{0.1}$Co$_{0.36}$Fe$_{0.36}$Mn$_{0.18}$ showed elevated Curie temperatures of 868~K and 1097~K, respectively, while also exhibiting substantially improved sustainability indices \cite {rowan2023cofemn0, li2024machine, chaudhary2021accelerated, bazioti2022probing}. These results clearly illustrate the utility of the mined database for application-specific and sustainability-aware alloy screening.

\subsubsection*{Corrosion resistant materials}

As the final case study, we focused on corrosion-resistant materials critical in applications such as marine structures, chemical processing equipment, energy infrastructure, transportation systems, and biomedical devices. 
Corrosion-resistant performance is typically characterized by low corrosion current density, reflecting reduced electrochemical activity and slower material loss under service conditions.
Again, for industrial benchmarking, 
we included materials spanning a wide range of aggressive service environments, including austenitic stainless steels (SS 304 and SS 316L) for general-purpose corrosion resistance in marine, chemical, and structural applications; \cite{loto2017study} nickel-based alloys such as Inconel 625 \cite{ganesan1991versatile} and Hastelloy C-276 \cite{lloyd2004cr} for highly corrosive and high-temperature environments encountered in chemical processing and energy infrastructure, titanium-based alloys including Ti-6Al-4V \cite{kumari2015laser} and Nitinol for biomedical, marine, and aerospace applications \cite{trepanier2000corrosion}, and high-alloy superaustenitic stainless steels such as AL-6XN for chloride-rich environments \cite{deverell1988corrosion}. 

This case study highlights an additional and important capability of the mined database. Corrosion current density is highly sensitive to the testing environment, and meaningful comparisons require strictly consistent exposure conditions. In contrast to datasets that report only property values, our database explicitly records the experimental testing conditions, enabling scientifically grounded and reproducible assessments of corrosion resistance. For this study, we therefore restricted the analysis to records measured in 3.5\% NaCl, a commonly used proxy for ambient and marine environments. Under this constraint, 342 alloy systems with both corrosion current density and sustainability indices known were identified, three of which can be seen to exhibit superior corrosion resistance while also achieving higher sustainability indices in Figure \ref{fig:Case_Studies}c. Two candidates, $\mathrm{Cr}_{0.18}\mathrm{Fe}_{0.62}\mathrm{Al}_{0.08}\mathrm{Ni}_{0.12}$ and $\mathrm{Cr}_{0.20}\mathrm{Fe}_{0.71}\mathrm{Mn}_{0.01}\mathrm{Ni}_{0.08}$, are compositionally very similar to SS 316L steel \cite{zhang2024corrosion, li2023fe61}, but the absence of Mo in these alloys leads to slightly higher sustainability indices while still achieving comparable or improved performance. Another promising candidate identified is the near-equiatomic alloy $\mathrm{Fe}_{0.27}\mathrm{Ni}_{0.25}\mathrm{Cr}_{0.24}\mathrm{Al}_{0.24}$, which exhibits a significantly lower corrosion current density \cite{zhang2024corrosion}. Analysis of elemental composition–property correlations indicates that higher Fe content is associated with increased corrosion current density (see Figure S2). Accordingly, the reduced Fe fraction in this alloy may contribute to its improved corrosion performance. Importantly, this type of analysis is not limited to the above mentioned conditions and can be readily extended to other standardized environments, such as 0.5~M H\(_2\)SO\(_4\) or 1~M HCl, enabling targeted materials selection under application-specific corrosion conditions.

While LLM-based data extraction shows significant promise for materials informatics and data-driven design, substantial challenges remain. Extending extraction capabilities beyond text and tables to include figures and other unstructured content is essential, particularly for records classified as out-of-scope in this work. Further, it is critical to capture microstructural information, especially for multiphase systems such as HEAs, as they significantly influence material properties.
Although some of this information is implicitly encoded through processing and characterization data, it is not yet sufficiently standardized to support rigorous, large-scale materials design. Advancing robust data standardization and representation will therefore be key to fully realizing the potential of LLM-driven materials discovery. With the rapid advancement in LLM capabilities and emergence of agentic frameworks, further progress in this area is expected.


\section{Conclusions}


In conclusion, this study presented a framework for the automated extraction of multi-element alloy data from scientific literature. The approach employed OpenAI’s conversational LLMs, GPT-4o and GPT-4o mini, together with task-specific prompt engineering to extract two complementary datasets: one derived from textual information in abstracts and experimental methods, and another obtained from tabular data, collectively spanning over 350 material properties. The resulting databases comprised of 37,711 and 148,069 records, respectively. To demonstrate the utility of resulting database, the extracted data was subsequently used to perform sustainable alloy selection across three application domains: lightweighting, soft magnetism, and corrosion resistance, identifying several tens of alloy compositions that outperform current industry benchmark materials in terms of both property performance and reduced environmental impact. To ensure accessibility and reproducibility, the datasets are made available through the \href{https://alloy.tattvasar.com/}{Alloy Tattvasar} website, and all source code, prompt templates, and few-shot examples are released on \href{https://github.com/MI-LAB-IITM/LLM_HEA_Data_Extraction/}{GitHub}. While the present work is limited to abstracts, experimental methods and tables, and omits information captured in figures or microstructures, the use of more advanced LLMs and agent-based frameworks provides a clear pathway for future extensions. The presented extraction pipeline is readily extensible to other classes of materials, and the resulting datasets are expected to enable data-driven approaches for sustainable materials design.

\section{Methods}

\subsection{Literature Corpus Construction and Pre-processing}

The literature corpus was restricted to textual and tabular content accessible via Elsevier API. The DOIs obtained from the Web of Science database were used to retrieve article content in structured XML format via the Elsevier API. For each publication, the XML files were accessed using an API key and institutional authentication token, and subsequently parsed using the \texttt{BeautifulSoup} library to extract hierarchical document elements, including titles, abstracts, section headings, and paragraph-level text.
The extracted XML content was transformed into comma-separated value (CSV) files, with paragraph-level text processed to appropriately handle embedded figure and table citations.
Tabular data were extracted using a modified version of the Table Extractor tool from the Olivetti group’s \href{https://github.com/olivettigroup/table_extractor}{repository}\cite{jensen2019machine}, which parses XML table structures to capture captions, footnotes, merged cells, and cell-level content, and converts each table into an individual CSV file. Metadata associated with the extracted tables were aggregated into a separate metadata file. This end-to-end pipeline enables structured, reproducible, and scalable data extraction from the literature, while respecting API rate limits and robustly handling common exceptions such as incomplete or missing XML content (see \href{https://github.com/MI-LAB-IITM/LLM_HEA_Data_Extraction/}{GitHub}).


\subsection{Prompts Used for the Data Extraction Task}

For QS1, four distinct extraction tasks were performed: alloy compositions, processing conditions, characterization techniques, and properties. A shared system prompt was used across all tasks to constrain model behavior, define scope, and enforce deterministic output formatting, while task-specific prompt templates incorporated domain-relevant universal information and curated few-shot examples. The universal information section was designed to span a wide range of reporting styles encountered in the literature, covering both fundamental concepts (e.g., stoichiometric notation) and more advanced or implicit representations of target information. The detailed example prompt for alloy composition extraction is provided below and illustrates the breadth of this contextual design. 
Few-shot examples were curated separately for each task and adapted according to target paragraph-level content. These examples were developed through preliminary experiments on 15 research articles (distinct from review articles used as evaluation set), which were also used to benchmark different LLMs (Table S1).
These were manually curated to explicitly address failure cases encountered in Table S1 and to span a diverse set of challenging scenarios.


For QS2, the extraction process was decomposed into two sequential LLM calls. In the first stage, the model was tasked solely with identifying table cells that correspond to material property headers. This stage employed a system prompt (included below) together with a user query containing the full table content, including the title and footer. The output of this step was a comma-separated list of unique table cells that contained property headers. The identified cells were subsequently used within a modified RAG framework to construct the context for the second stage. Specifically, the RAG component retrieved a set of likely material properties associated with the detected headers, which were then incorporated into the context section of the second prompt. This context included concise definitions of each property, common symbols and notations, typical reporting formats, and frequently used units as encountered in the literature. This contextual information for each property was curated offline using GPT-4o, and was subsequently manually reviewed and corrected. In total, this process was applied to 354 material properties. The full template for the second stage prompt is provided below. See \href{https://github.com/MI-LAB-IITM/LLM_HEA_Data_Extraction/}{GitHub} for the complete list curated examples, article DOIs and source code used for data extraction.

\begin{tcolorbox}[
    title=QS1: Prompt Template for Alloy Composition Extraction,
    fonttitle=\bfseries,
    colback=white,
    colframe=gray!40,
    colbacktitle=gray!80,
    coltitle=white,
    boxrule=1.5pt,
    arc=4mm,
    left=4mm,
    right=4mm,
    top=5mm,
    bottom=4mm,
    fontupper=\footnotesize,
    halign=left
]
\setlength{\parskip}{1pt}
\setlength{\itemsep}{1pt}
\setlength{\topsep}{1pt}

\textbf{System Prompt:}
\begin{itemize}[noitemsep, topsep=0pt]
    \item You are an assistant designed to extract materials data from scientific articles.
    \item Take your time, but be accurate in your responses.
    \item You reply in a single word or a comma separated list of words.
    \item Evaluate any mathematical expression in your final output.
\end{itemize}

\medskip
\textbf{Instructions:}
\begin{itemize}[noitemsep, topsep=0pt]
    \item Keep the response to just one word.
    \item If you don't have any context and are unsure of the answer, reply with `---' as your response.
    \item If more than one answer is present, name all of them separated by comma.
    \item Strictly follow the three steps below.
\end{itemize}

\medskip
\noindent\textbf{Step 1:} Read the information below (universal data applying to all queries). Answer the user query regarding the information given in the article.

\medskip

\noindent\textbf{Universal Information:}
\begin{enumerate}
\item An alloy is a solid mixture of multiple elements. Its chemical composition conveys the relative amount of each element present. For example, an alloy with chemical composition AlCoCr contains Al, Co, and Cr in equal proportions. Similarly, an alloy with composition Al\textsubscript{2}Co\textsubscript{3}Ni\textsubscript{5} contains 2 parts Al, 3 parts Co, and 5 parts Ni.

\item Alloy compositions may be explicitly reported in the text. For example, in the sentence: ``An alloy system was synthesized through arc melting the constituents at 1500 K to obtain Al\textsubscript{2}Mo\textsubscript{4}Cr\textsubscript{6} and AlMo\textsubscript{3}Cr\textsubscript{5},'' the alloy compositions mentioned are Al\textsubscript{2}Mo\textsubscript{4}Cr\textsubscript{6} and AlMo\textsubscript{3}Cr\textsubscript{5}.

\item Alloy compositions may also be represented implicitly using molar ratios or atomic ratios instead of explicit stoichiometric formulas.

\item In the molar ratio notation, variable element content is specified using a parameter. For example, ``Multicomponent AlCoCrCuFeNiMo\textsubscript{x} (x values in molar ratio, $x = 0$, $0.2$, $0.8$, and $1.0$) alloys were prepared using copper mould casting.'' This corresponds to four alloy compositions: AlCoCrCuFeNi, AlCoCrCuFeNiMo\textsubscript{0.2}, AlCoCrCuFeNiMo\textsubscript{0.8}, and AlCoCrCuFeNiMo.

\item A common convention for reporting alloys with varying elemental percentages uses symbolic placeholders. For example, Al\textsubscript{x}TiVCr ($x = 1, 2$) refers to two alloys: AlTiVCr for $x = 1$ and Al\textsubscript{2}TiVCr for $x = 2$.

\item In some cases, compositions are reported using atomic percentages with grouped elements. For example, Al\textsubscript{x}(TiVCr)\textsubscript{100--$x$} ($x = 25$, $40$ at.\%) represents two alloys: Al\textsubscript{25}Ti\textsubscript{25}V\textsubscript{25}Cr\textsubscript{25} and Al\textsubscript{40}Ti\textsubscript{20}V\textsubscript{20}Cr\textsubscript{20}. Here, Al contributes $x$\% of the composition, while the remaining $(100 - x)$\% is equally distributed among the elements inside the brackets.

\item For bracketed expressions, the composition is evaluated in two steps: first, compute the remaining fraction $(100 - x)$; second, divide this fraction equally among the elements within the brackets. For example, when $x = 25$, $(100 - x) = 75$\%, which is divided equally among Ti, V, and Cr, yielding 25\% each and resulting in Al\textsubscript{25}Ti\textsubscript{25}V\textsubscript{25}Cr\textsubscript{25}.
\end{enumerate}

\medskip
\noindent\textbf{Step 2:} Go through these example prompts for extracting alloy composition:

\begin{description}[itemsep=2pt, labelwidth=*, leftmargin=!, font=\normalfont\bfseries]
    \item[Example 1:]
    \textit{Article:} It is found that both of these aforementioned alloys have hardnesses of about 420 HV, which is equated to their similar microstructures.\\
    \textit{Answer:} ---

    \item[Example 2:]
    \textit{Article:} A multi-phase region can have multiple crystal structures like FCC+BCC or BCC+HCP etc. The as-cast AlCoCrCuNi base alloy is found to have a dendritic structure, of which only solid solution FCC and BCC phases can be observed.\\
    \textit{Answer:} AlCoCrCuNi

    \item[Example 3:]
    \textit{Article:} AlCrFeCoNiCuTi alloy contains BCC1 phase, BCC2 phase and a FCC phase. AlCrFeCoNiCuV alloy contains two different phases with BCC, FCC structure, respectively.\\
    \textit{Answer:} AlCrFeCoNiCuTi, AlCrFeCoNiCuV

    \item[Example 4:]
    \textit{Article:} Al$_{x}$Ti (x=1, x=2, x=3) are a class of alloys used for various applications.\\
    \textit{Answer:} AlTi, Al$_{2}$Ti, Al$_{3}$Ti
\end{description}

\medskip
\noindent\textbf{Step 3:}
\noindent\textbf{User Question:} Follow the examples, context and to extract the alloy compositions mentioned in the article.\\
\textbf{Article:} \{article\}\\
\textbf{Answer:}
\end{tcolorbox}



\begin{tcolorbox}[
    title=QS2: System Prompt for Identifying Property Header Cells,
    fonttitle=\bfseries,
    colback=white,
    colframe=gray!40,
    colbacktitle=gray!80,
    coltitle=white,
    boxrule=1.5pt,
    arc=4mm,
    left=4mm,
    right=4mm,
    top=5mm,
    bottom=4mm,
    fontupper=\footnotesize,
    halign=left
]
\setlength{\parskip}{3pt}

You are a materials science expert identifying cells in a table that have material property names. The
table, title, and footer too may contain property name include them too. You must respond with cells
that you think contain a property name in a comma separated format. Make sure each of the properties in your response mentioned are unique.
\end{tcolorbox}

\begin{tcolorbox}[
    title=QS2: Prompt Template for Material Property Extraction From Tables,
    fonttitle=\bfseries,
    colback=white,
    colframe=gray!40,
    colbacktitle=gray!80,
    coltitle=white,
    boxrule=1.5pt,
    arc=4mm,
    left=4mm,
    right=4mm,
    top=5mm,
    bottom=4mm,
    fontupper=\footnotesize,
    halign=left
]
\setlength{\parskip}{1pt}
\setlength{\itemsep}{1pt}
\setlength{\topsep}{1pt}

\textbf{System Prompt:}
\begin{itemize}
\item You are a materials science expert extracting property values available in tables in materials literature.
\item You must respond in a comma separated format, containing 6 columns: alloy compositions, processing conditions, testing conditions, property, value and unit.
\item Take your time, but be accurate in your responses.
\item Evaluate any mathematical expression in your final output.
\end{itemize}

\textbf{Instructions:}
\begin{itemize}
\setlength\itemsep{0pt}
\setlength\parskip{0pt}
\setlength\parsep{0pt}
\item Identify properties from context section described in the table.
\item If no properties found, reply with `-'.
\item Return data in CSV format with columns: Alloy, Processing condition, Testing condition, Property, Value, Unit.
\item Extract only properties available in context; ignore unfamiliar headings.
\item Include testing conditions (strain rates, temperature) as noted in table.
\end{itemize}

\noindent\textbf{Context (Material Properties):}

\textbf{1. Ultimate Tensile Strength (UTS)}:\\
\textbf{Definition}: UTS is the maximum stress a material can withstand before breaking, measured during a tensile test. The test can be done under tension or compression, unless mentioned explicitly assume that the test is done under tension. Report the property as Ultimate Tensile Strength(Tension) if done under tension else report Ultimate Tensile Strength(Compression). Testing conditions mentioned are usually the strain rates and temperature. \\
\textbf{Units:} Pascals (Pa), megapascals (MPa), or pounds per square inch (psi). \\
\textbf{Comparison:} UTS is always higher than yield strength since it represents the material's maximum capacity to resist tension before fracture. \\
Variations: Often called tensile strength or breaking strength.

...

\textbf{n. Hardness:} \\
...

\textbf{n+1. Other Properties:} There are a lot of properties with clear commonly used definitions, like Density, Melting point, Boiling point etc. In addition to the n properties mentioned above, if you are sure that one of the following properties is present, extract them in the prescribed format.

\textbf{n+2. Testing Conditions:} Properties for a material cannot always be described through a number alone. For example, the Young's modulus of a material can vary with temperature and this is usually mentioned in the table details; be sure to extract this and report it as a testing condition. 

\textbf{n+3. Note:} Some properties like refractive index, Poisson's ratio do not have units, report units as "-".

\noindent\textbf{Example 1:}

\noindent\textit{Table Caption:} Table 4. Hardness of Al\textsubscript{0.5}CoCrCuFeNi alloys in different states.

\noindent\textit{Table Data:}
\begin{center}
\small
\begin{tabular}{|c|c|}
\hline
\textbf{State} & \textbf{Hardness (HV)} \\
\hline
As-cast & 208±2 \\
As-forged & 350±3 \\
\hline
\end{tabular}
\end{center}

\noindent\textit{Output:}
\begin{center}
\small
\begin{tabular}{|c|c|c|c|c|c|}
\hline
\textbf{Alloy} & \textbf{Proc. Cond.} & \textbf{Test. Cond.} & \textbf{Property} & \textbf{Value} & \textbf{Unit} \\
\hline
Al\textsubscript{0.5}CoCrCuFeNi & As-cast & --- & Hardness & 208±2 & HV \\
Al\textsubscript{0.5}CoCrCuFeNi & As-forged & --- & Hardness & 350±3 & HV \\
\hline
\end{tabular}
\end{center}

...

\noindent\textbf{Example 4:} 
...

\noindent\textbf{User Task:} Extract material properties from the provided table following the context and instructions above. Return results in CSV format.

\noindent\textbf{Table Caption:} \{caption\}

\noindent\textbf{Table Data:} \{table\_data\}

\noindent\textbf{Output:}
\end{tcolorbox}


\subsection{DB2 Post-processing}

To use the database for downstream analysis, it is important to have exact chemical compositions. Here, standardized alloy entries are defined as those entries that have chemically valid compositions comprising solely of element symbols and physically meaningful stoichiometric values. This representation is necessary to enable composition-based analysis, materials selection, and sustainability studies. A significant fraction of the extracted alloy records employed non-standard nomenclature, including laboratory shorthand, acronyms, trade names, and article-specific aliases whose definitions were often absent in the target table but included elsewhere in the manuscript. To standardize entries in the database, we implemented a two-stage post-processing pipeline to standardize alloy compositions, first using a rule-based check on the compositions and the next employing another LLM call with context of the non-standard terminology.


In the initial QS2 extraction, certain records listed exact chemical compositions as property values and used non-standard acronyms or identifiers as alloy names. As part of a rule-based validation step, the extracted compositions were therefore used to replace non-standard alloy identifiers across all records from the same article.
Cases that remained ambiguous after this step were resolved using an LLM. For each such identifier, the context surrounding its first occurrence in the paper was provided to the model, along with a small set of curated examples retrieved via a RAG framework, as shown below. Nearly 2,400 identifiers were standardized this way and corresponded to over 10,000 records in DB2. These LLM-standardized entries were manually validated before updating DB2.

\begin{tcolorbox}[
    title=Prompt Template for Alloy Composition Standardization,
    fonttitle=\bfseries,
    colback=white,
    colframe=gray!40,
    colbacktitle=gray!80,
    coltitle=white,
    boxrule=1.5pt,
    arc=4mm,
    left=4mm,
    right=4mm,
    top=5mm,
    bottom=4mm,
    fontupper=\footnotesize,
    halign=left
]
\setlength{\parskip}{2pt}

\textbf{Task: Alloy Composition Standardization}

A block of text where each line contains a potentially unconventional alloy name or composition
formula. These may include typographical variations, shorthand forms, acronyms, or contextual
abbreviations (e.g., ``SS304'', ``Ti6Al4V'', ``Ni-base superalloy'', ``Fe--Cr--Ni--Mo steel'').

\medskip
\noindent\textbf{User Input}

\begin{enumerate}[leftmargin=*, noitemsep]
    \item \textbf{Entries to Map:}  
    A list of potentially unconventional alloy identifiers to be standardized.
    \item \textbf{Context:}  
    Text extracted from the associated research article, including:
    \begin{itemize}[noitemsep, topsep=0pt]
        \item Title
        \item Abstract
        \item Relevant extracted paragraph(s)
    \end{itemize}
    \item \textbf{Few-Shot Examples:}  
    Curated examples of challenging alloy name conversions identified during pilot analysis.
\end{enumerate}

\medskip
\noindent\textbf{Instructions}
\begin{itemize}[noitemsep, topsep=0pt]
    \item Identify whether each entry can be mapped to a chemical composition using only the
    provided context.
    \item Convert identifiable compositions into standard metallurgical notation.
    \item Assign a confidence level (\textit{high}, \textit{medium}, or \textit{low}) to each mapping.
    \item If the composition cannot be determined unambiguously, return the original entry unchanged.
    \item Do not infer or introduce information not explicitly present in the context.
\end{itemize}

\medskip
\noindent\textbf{Standardization Guidelines}
\begin{itemize}[noitemsep, topsep=0pt]
    \item Use valid chemical element symbols (e.g., Al, Ti, Fe, Ni, Cr).
    \item Use standard composition notation (e.g., Al$_{85}$Ti$_{15}$, CoCrFeNiMo$_{0.2}$).
    \item Hyphens typically indicate alloy families rather than compositions and should be treated with caution.
    \item Substitute variables (e.g., $x$) only if their values are explicitly defined in the context.
    \item Preserve bracketed substructures without modification.
\end{itemize}

\medskip
\noindent\textbf{Expected Output Format}
\begin{verbatim}
[
  {
    "original": "entry_to_map_1",
    "standardized": "standard_composition_1",
    "confidence": "high"
  },
  {
    "original": "entry_to_map_2",
    "standardized": "entry_to_map_2",
    "confidence": "low"
  }
]
\end{verbatim}

\end{tcolorbox}
\medskip

\section*{Author Contributions}
A.K.S and R.B. conceived the project.
S.M.K.D contributed to development of pre-processing pipeline to download and save articles in a structured format.
A.K.S developed the QS1 and QS2 pipeline.
S.M.K.D developed code for large-scale dataset extraction.
M.C. post-processed the DB2. B.P.P helped with LLM evaluation and validation.
All the authors contributed to the data analysis and to the preparation of the manuscript. A.K.S and R.B. wrote the manuscript with input from all the co-authors. All authors read and approved the manuscript. R.B. supervised and directed the overall project.

\section*{Acknowledgements} \par 
R.B. acknowledges ANRF (File No. SRG/2023/001055), DRDO-DIA and WSAI for funding. Computational resources
from Robert Bosch Centre for Data Science and AI, IIT Madras, are also acknowledged. The authors thank Shreya Kumari, Shivangi Keshri and Prof. Satyesh Yadav for their assistance in deployment of database on Alloy Tattvasar.

\section*{Conflicts of Interest}
The authors declare no conflict of interest.

\section*{Data Availability Statement}
The LLM-extracted dataset can be accessed via \href{https://alloy.tattvasar.com}{Alloy Tattvasar} webpage, while the accompanying LLM source code is provided on \href{https://github.com/MI-LAB-IITM/LLM_HEA_Data_Extraction/}{GitHub}. Any other data that support the findings of this study are available from the corresponding author upon reasonable request.

\medskip

%

\bibliographystyle{unsrt}
\bibliography{references}

\end{document}